\documentstyle[11pt,paspconf]{article}

\begin{document}

\def\half{{\scriptstyle {1 \over 2}}}
\def\ie{{\it {\frenchspacing i.{\thinspace}e. }}}
\def\eg{{\frenchspacing e.{\thinspace}g. }}
\def\cf{{\frenchspacing\it cf. }}
\def\etal{{\frenchspacing\it et al.}}
\def\simlt{\hbox{ \rlap{\raise 0.425ex\hbox{$<$}}\lower 0.65ex\hbox{$\sim$} }}
\def\simgt{\hbox{ \rlap{\raise 0.425ex\hbox{$>$}}\lower 0.65ex\hbox{$\sim$} }}
\def\solar{\odot}
\def\msun{{\rm M}_\odot}
\def\rsun{{\rm R}_\odot}
\def\pc{{\rm pc}}
\def\Myr{{\rm Myr}}
\def\Gyr{{\rm Gyr}}
\def\Rf{\parindent=0pt\medskip\hangindent=3pc\hangafter=1}

\title{The Role of Binaries in the Dynamical Evolution of Globular Clusters}
\author{Piet Hut}
\affil{Institute for Advanced Study, Princeton, NJ 08540, U.S.A.}

\begin{abstract}

Three important developments are vastly increasing our understanding
of the role of binaries in the dynamical evolution of globular
clusters.  From the observational side, the Hubble Space Telescope has
shown us detailed pictures of the densest areas in post-collapse
cluster cores.  From the computational side, the Grape-4
special-purpose hardware is now allowing us to model small globular
clusters on a star-by-star basis, and has already given us the first
direct evidence of the occurrence of gravothermal oscillations in such
systems.  From the theoretical astrophysics side, integrated
simulations are now becoming feasible that combine stellar dynamics
with stellar evolution and hydrodynamics.  Given these three
developments, we can expect the current rapid progress in our
understanding of globular cluster evolution to continue at an even
higher rate during the foreseeable future.  In this review an outline
is given of the current status of globular cluster simulations, and
the expected progress over the next five years.

\end{abstract}

\section{Introduction}

Star cluster dynamics is like a ballet with about a million
performers, but it is the binary stars which take center stage, for it
is in their {\it pas de deux} that most of the action develops.  And
it is through their interactions with other players, in the form of
occasional {\it pas de trois} and {\it pas de quatre}, that
fascinating new patterns (and even new characters) arise.

In this conference, we have already heard several talks in which the
dynamics of binary stars has been reviewed, from various angles.
Therefore, instead of attempting a full-fledged review, I am happy to
refer to the presentations by Aarseth, Clarke, Leonard, Mardling,
McMillan, Phinney and Rasio, all of whom discuss aspects of the
question of how to model interactions of binary stars with their
surroundings.  In addition, many references to related work can be
found in the review article by Hut {\it et al.} (1992).  Given all
these pointers, I will take the liberty to focus on the future more
than on the past, sketching the expected developments during the next
five to ten years.

In this contribution, I will first address the nature of the dynamical
evolution of star clusters, in \S2.  Then, in the point-mass
approximation, I will sketch expected developments in the purely
gravitational $N$-body system, in \S3.  With the addition of effects
of stellar evolution and the hydrodynamics of stellar encounters, more
realistic modeling efforts are previewed in \S4.  Conclusions and a
general outlook are presented in \S5.

\section{Ecology}

Nearly everywhere in our galaxy, the local dynamics of multiple star
systems and the global dynamics of the galaxy as a whole can be neatly
separated.  In the solar neighborhood, a double star (or triple or
other multiple star system) with a diameter of a few AU will have a
negligible chance to interact with neighboring stars, even on a time
scale comparable to the current age of the galaxy.  A rough estimate
suffices here.  With a density $n$ of one star per cubic parsec, a
relative velocity $v$ of 50 km/s, and a target area $\sigma$ with a
radius $r$ of 5 AU, the rate of close encounters is $n\sigma v = \pi
r^2 v n = 10^{-13}/yr$.  This gives a probability of only
0.1\% for such an event to occur during the next Hubble time.

More distant encounters do occur of course, but a single encounter is
relatively harmless, and it typically takes many encounters to unbind
a wide binary ({\it cf.} Hut 1985 for a graphic illustration).  As a
result, only the widest binaries in the solar neighborhood, with
separations of order $10^3-10^4$AU, are potentially subject to
dissolution by encounters with passing stars and molecular clouds.
However, even in this case the local-global interaction is a one-way
street: the effect of the environment on the local system can be
dramatic, but the feedback effects on the environment are negligible.
A binary with a separation of $10^3$ AU typically has a binding energy
that is three orders of magnitude lower than the kinetic energy
associated with the relative motion of single stars in the solar
neighborhood.  Whether or not such a binary is dissolved hardly makes
a mark on the motion of the field stars.  Clearly, the dynamical
interactions with binaries and multiple star systems form an utterly
negligible component in the energy budget of the galaxy as a whole.

Even in the densest parts of the galaxy, in the inner parsec of the
galactic nucleus, the energy locked up in the internal motions of
binaries is likely to be at least an order of magnitude less than the
energy available in the motions of the single stars and the
center-of-mass motions of the binaries and higher multiple star
systems.  With a velocity dispersion of more than 100 km/s, only the
tightest binaries will have an orbital speed exceeding the typical
center-of-mass motion, and these binaries will only contain a small
fraction of all the stars in the field.

The situation is just the opposite in the case of star clusters.  Both
open clusters and globular clusters have a much lower velocity
dispersion than the galactic center, and a much higher density than
the solar neighborhood.  The combined effect gives a situation in
which a typical binary can easily have an orbital velocity far
exceeding the velocity dispersion of the cluster, and therefore have
an energy $\gg 1$kT.  Combined with the fact that observations show us
that a fair fraction of the stars in clusters have been formed as
binaries ({\it cf.} the talks during the first day of this conference,
in the session `The Search for Duplicity'), it is clear that binaries
play an important role in the dynamics of star clusters.  

As a result, the total energy locked up in binary binding energy is at
least comparable to, and in some cases may well exceed, the total
energy of the cluster as a whole (in the form of the kinetic and
potential energy of the single stars and of the centers of mass of the
binaries).  Given this situation, changes in binary properties that
take place during the course of normal stellar evolution will have a
repercussion on the dynamical evolution of a star cluster as a whole.
In addition, close encounters involving a combination of single stars
and binaries can affect the parameters of the binaries in very complex
ways.  Either type of process, internal evolution in relatively
isolated binaries, or three-body and four-body encounters, will modify
the balance between the two energy budgets of a cluster, governing the
external and internal degrees of freedom (bulk energy and total binary
binding energy, respectively).

While it is the macroscopic energy budget that drives the dynamical
evolution of a star cluster as a whole, this budget can be
significantly modified through the strong coupling with the comparable
microscopic energy budget of internal degrees of freedom of binary
stars.  The feedback mechanisms between stellar dynamics and stellar
evolution therefore play a vital role in the evolution of star
clusters.  The term `ecology', introduced by Heggie (1992), captures
the essence of this interplay.

\section{The Gravitational $N$-Body System}

Even without the complexities of mass overflow and the possibility of
physical collisions in interactions between stars, the dynamical
evolution of a star cluster already poses a daunting problem on the
point-mass level.  At the time of the conference, routine calculations
of star cluster evolution typically featured a total number of particles
in the range of $1-5\times10^3$, with the record being held by Spurzem
\& Aarseth (1996) for a run with $10^4$ particles.

Why is it so hard to model a globular cluster with a more realistic
number of particles, say $10^5-10^6$?  After all, in cosmological
simulations such numbers are routine these days.  The problem lies in
the enormous discrepancy of length and time scales in the dynamics of
globular clusters.  The size of a main sequence star is a factor $10^9$ smaller
than the size of a typical cluster.  If neutron stars are taken into
account, the problem is worse, and we have a factor of $10^{14}$
instead, for the discrepancy in length scales.  The time scale on
which clusters evolve, of order ten billion years, gives a discrepancy
of time scales of a factor $10^{14}$ for Kepler orbits close to normal
stars, or $10^{21}$ for neutron stars.  As a result, globular cluster
simulations are rather delicate affairs, and it is far from easy to
get a code to work at all, bridging these vast scales, let alone to
get a code to work efficiently ({\it cf.} Aarseth's contribution in
these proceedings).

Currently, with routine type calculations, it is only feasible to
model the evolution of a globular cluster containing roughly
$N=5\times10^3$ stars, since this requires some $10^{15}$ floating
point calculations, equivalent to 10 Gflops-day, or a year or more on
a typical workstation.  The cpu cost scales $\propto N^3$, where the
inter-particle forces contribute two powers in $N$ and the increased
time scale for heat conduction contributes the third factor of $N$.
Therefore, a calculation with half a million stars, resembling a
typical globular star cluster, will require $\sim 10$ Pflops-day (see
Makino \& Hut 1988, 1990 for more accurate scaling estimates).

While Pflops speeds are out of the question at present, it is
important to note that it is only speed that is lacking now, not
memory.  In contrast to most other types of compute-hungry future
calculations, the memory requirements for modeling a complete globular
cluster are relatively modest.  All that is needed is to keep track of
$N=5\times10^5$ particles, each with a mass, position, velocity, and a
few higher derivatives for higher-order integration algorithms.
Adding a few extra diagnostics per particle still will keep the total
number of words per particle to about 25 or so.  With 200 bytes per
particle, the total core memory requirement will be a mere 100 Mbytes.

Output requirements will not be severe either.  A snapshot of the
positions and velocities of all particles will only take 10 Mbytes.
With, say, $10^5$ snapshot outputs for a run, the total run worth 10
Pflops-day will result in an output of only 1 Tbyte, far less than
what will be required by typical hydrodynamical calculations that
will be carried out on a Pflops machine.

It was through these kind of considerations that a group of
astrophysicists at Tokyo University decided in 1989 to begin building
a series of special-purpose pieces of hardware (Sugimoto {\it et al.}
1990).  The basic idea was to use a workstation to integrate particle
orbits, while leaving the computation of the gravitational forces
between the particles to the special hardware.  Various versions of
this GRAPE hardware have been build, and have been made available at
modest cost to outside researchers as well.  Price/performance ratios
for these machines are far smaller than for commercially available
computers, by orders of magnitude.

The name GRAPE stands for GRAvity PipE, and indicates a family of
pipeline processors that contain chips specially designed to calculate
the Newtonian gravitational force between particles.  A GRAPE
processor operates in cooperation with a general-purpose host
computer, typically a normal workstation.  Just as a floating point
accelerator can improve the floating point speed of a personal
computer, without any need to modify the software on that computer, so
the GRAPE chips act as a form of Newtonian accelerator.  The force
integration and particle pushing are all done on the host computer,
and only the inter-particle force calculations are done on the GRAPE.
Since the latter require a computer processing power that scales with
$N^2$, while the former only require $\propto N$ computer power, load
balance can always be achieved by choosing $N$ values large enough.

A significant step toward the modeling of globular star clusters has
been made in the summer of 1995 with the completion of the GRAPE-4,
the fastest machine to date, with a speed of more than 1 Tflops.  It
can be driven reasonably efficiently by a workstation of 100 Mflops.
Although such a workstation operates at a speed that is lower than
that of the GRAPE by a factor of $10^4$, a fair fraction of the
Teraflops peak speed can be reached, allowing simulations of up to
$10^5$ particles to reach core collapse and beyond, in the case of
unequal masses (for equal masses, the maximum number of particles is
somewhat lower, since those calculations are more compute intensive).
The first scientific results of the GRAPE-4, including the first
convincing evidence of gravothermal oscillations in $N$-body
simulations, predicted by Sugimoto \& Bettwieser (1983), have been
presented at the I.A.U. Symposium 174 in Tokyo, in August 1995 by
Makino (1996).

The next, and definitive step that will enable any globular cluster to
be modeled realistically might take place as early as the year 2000.
If funding can be found, there is no technological obstacle standing
in the way of a speedup of the current GRAPE-4 machine by a factor of
a thousand, during the next five years.  Most of this speed-up will
come from further miniaturization, allowing a larger number of gates to
be mounted on a single chip, and allowing a higher clock speed as
well.

A Petaflops machine by the year 2000, allowing simulations of core
collapse and post-collapse evolution of up to $10^6$ particles, is
thus a plausible goal.  This would remove any hardware barrier towards
realistic simulations of globular clusters.  There would be two
barriers remaining.  One is the formidable task to integrate our
current knowledge of hydrodynamics and stellar evolution with such
large-scale stellar dynamical simulations.  The other barrier concerns
our current lack of knowledge of the physics of some of the processes
involved in multiple-star evolution, such as common-envelope
evolution.

\section{The Astrophysical $N$-Star System}

The simplest way to model a star cluster on a star-by-star basis is to
approximate each star as a point mass, starting with an equal-mass
system and keeping those stellar masses constant and equal throughout
the calculation.  While this is clearly a rather unrealistic
approximation, it does allow us to model the overall behavior of a
cluster, namely its core collapse and post-collapse expansion.  These
two phenomena are both regulated by the heat conduction provided by
two-body relaxation, a process that does not require close encounters
to be effective.  Heat conduction, therefore, can be well described in
the point-mass approximation.  Heat production, however, takes place
in and near the core in the post-collapse phase, and an accurate
modeling of this heat engine requires a careful investigation of the
physics of close binary stars, well beyond the point-mass
approximation.

But even in pre-collapse calculations, the point-mass approximation
quickly becomes unrealistic, as soon as we drop any of the
restrictions from the simplest equal-mass calculation.  For example,
the inclusion of a mass spectrum in a $N$-body simulation will lead to
mass segregation.  In a real star cluster, too, most of the heavier
stars tend to sink to the central area on a half-mass relaxation time
scale.  However, once they have arrived in the center, they do not
remain there forever, since they are the first stars to burn out and
loose considerable amounts of mass in their transitions to the final
stage of stellar evolution.  For most stars, this final stage is a
white dwarf, with a mass well below that of the main sequence
turn-off, which means that such a star will be expelled from the core
that was dominated by its progenitors.

In contrast to the real world, a simple-minded inclusion of a mass
spectrum in a $N$-body simulation will feature a indefinite piling-up
of massive stars in the center.  Of course, it is possible to repair
this unrealistic situation by given the particles a finite lifetime,
after which they shed a specified amount of mass, depending on their
original mass.  This type of recipe is the first step towards
providing a simple form of stellar evolution for point particles.
However, such a recipe does not specify what will happen in the case
of a binary star.

For example, if two heavy core stars will form a binary in a
relatively tight orbit, with a pericenter distance significantly
smaller than 1 AU, it will not be possible for those stars to evolve
in isolation.  Instead, the heavier of the two will undergo mass
overflow onto the companion star, in a process that may or may not
lead to a spiral-in and a common-envelope phase that could lead to a
coalescence of the two stars.  And it is precisely this type of hard
binary, with a binding energy much larger than the kinetic energy of a
typical single star, that plays an important role in the energy
generation in the core, especially in the post-collapse phase.  If the
simplest stellar evolution recipes cannot handle the complexities of
binary stars, there is not much point in using them at all.

The conclusion is a sobering one: as soon as we want to make an
equal-mass point particle $N$-body simulation more realistic, we are
immediately driven to add three successive refinements, namely: to
include a mass spectrum; therefore to give a prescription for burn-out
at finite ages; and consequently to give much more complex recipes for
how binaries behave --- not only in isolation, which is already a
complex problem, but also during interactions in which three or more
stars are involved simultaneously and dynamically.

It is not surprising that such recipes have not yet been developed to
a mature stage.  While finding a reasonable answer for most of the
cases of interest is not too difficult, such recipes are not very
useful if they give answers only in, say, 99\% of the cases.  A
cluster simulation with $10^4$ stars, and a primordial binary fraction
of 10\% or more, will feature many thousands of strong three-body and
four-body encounters, in which collisions may take place and sudden
mass transfer may be initiated.  It will not be practical to have to
stop the simulation, even for only 1\% of those encounters, in order
to study the situation in detail and give an {\it ad hoc}
prescription.

On the other hand, a fully foolproof set of recipes is unlikely to
emerge from pure thought alone.  For example, in some cases two
different mass-exchanging binaries will meet each other, and it will
be quite a challenge to provide a fully automated prescription for
what will happen in such a case.  Most likely, it will require a
significant amount of experimentation in the form of realistic
simulations, before a reliable suite of recipes will be developed.

Several groups are currently developing such sets of recipes.  One
approach is to start directly with $N$-body simulations, adding
stellar evolution prescriptions to close encounter situations.  A
report of such a method can be found in these proceedings, in the
contribution by Aarseth.

Another approach is to model the core of a dense star cluster as a
homogeneous stellar system, leaving out at first two-body relaxation
effects and the overall dynamical cluster evolution, while
concentrating solely on the effect of close encounters.  Using the
automated three-body scattering package described by McMillan in his
contribution to these proceedings, Portegies Zwart has begun to
implement a complete set of recipes for stellar collisions as well as
stellar evolution effects, in order to model the evolution of an
otherwise static cluster core ({\it cf.} his talk at this symposium,
and Hut 1995, where an example is given of one of his calculations in
the form of an interplay between three stars, leading to the formation
of two blue stragglers).

\section{Conclusions and Outlook}

Fifteen years ago, fundamental aspects of the dynamical evolution of
globular clusters were still unknown.  The results of core collapse, as
well as the character of the subsequent evolution, were hardly
explored.

Ten years ago, several approximate treatments, such as Fokker-Planck
models and conducting gas sphere models, had begun to explore the
nature of post-collapse evolution, and had shown the ubiquitous
presence of a new and unexpected phenomenon: gravothermal
oscillations, a form of thermodynamic instability in the inner few
percent of the cluster mass.

Five years ago, another major new ingredient had become common-place
in cluster simulations: the presence in globular clusters of a
significant fraction of primordial binaries, something that greatly
affected the dynamics of the post-collapse evolution, suppressing core
oscillations as long as most of these primordial binaries had not yet
been destroyed or ejected.

Around the time of this conference, full $N$-body calculations have
begun to approach the size of small globular clusters, and have
already delivered hard evidence for the existence of gravothermal
oscillations in realistic cluster models.  At the same time, realistic
simulations are starting up that include not only gravitational
stellar dynamics, but also hydrodynamical effects of collisions and
close encounters, as well as a host of stellar evolution effects.

What can we expect the state of the art to be around the year 2000,
another five years from now?  On the hardware side, if we are lucky,
we may be able to extend our $N$-body calculations by another factor
of ten in $N$, which will enable us to simulate all types of globular
clusters on a star-by-star basis.

On the software side, we can expect to have several suites of stellar
evolution recipes, integrated with stellar dynamical codes, which will
allow us to make realistic simulations that include a large range of
non-gravitational astrophysical effects.  Perhaps some of these
integrated systems will even have small and robust stellar evolution
codes running in parallel with the background stellar dynamics code,
in order to solve complex situations from scratch.

Finally, on the side of our understanding of fundamental physical
processes, we can express the hope that current bottlenecks, such as
uncertainties concerning common envelope evolution, will slowly begin
to yield their secrets, as a result of ever more detailed
three-dimensional hydrodynamical simulations.  However, the full
resolution of some of these bottlenecks is expected to take far
longer than five years.

\bigskip
\acknowledgments
I thank Douglas Heggie for helpful comments on the manuscript.

\end{document}